# Synchronization Instability of Inverter-Based Generation During Asymmetrical Grid Faults

Xiuqiang He, *Student Member, IEEE*, Changjun He, Sisi Pan,
Hua Geng, *Fellow, IEEE*, and Feng Liu, *Senior Member, IEEE*

*Abstract*—The transient stability of traditional power systems is concerned with the ability of generators to stay synchronized with the *positive-sequence* voltage of the network, whether for symmetrical or asymmetrical faults. In contrast, both *positive- and negative-sequence* synchronizations should be of concern for inverter-based generation (IBG) under asymmetrical faults. This is because the latest grid codes stipulate that IBG should inject dual-sequence current when riding through asymmetrical faults. Currently, much less is known about the synchronization stability during asymmetrical faults. This significantly differs from the positive-sequence synchronization alone because the coupled dual-sequence synchronization is involved. This paper aims to fill this gap. Considering the sequence coupling under asymmetrical faults, the dual-sequence synchronization model of IBG is developed. Based on the model, the conditions that steady-state equilibrium points should follow are identified. The conditions throw light on the possible types of synchronization instability, including the positive-sequence dominated instability and the negative-sequence dominated one. For different types of instability, the dominant factors are analyzed quantitatively, which are reflected by the limit on the current injection amplitude. Exceeding the limit will lead to the loss of both positive- and negative-sequence synchronizations. The model and the analysis are verified by simulations and hardware-in-the-loop experiments.

## I. INTRODUCTION

RECENT years have witnessed a wide use of power electronic inverters as the interface of renewable energy generation in the power system. Remaining in synchronism with the grid frequency is the foundation for the stable operation of inverter-based generation (IBG) [1]. Different from the intrinsic synchronization mechanism of synchronous generators, the synchronization of IBG is mainly accomplished by detecting and following the phase-angle of the terminal voltage. As the share of IBG is continuously rising in the power system, such a grid-following mechanism has become one of the basic principles governing the synchronization behavior of the power system. For the IBG connected to a weak grid, in particular under grid faults, the terminal voltage is much sensitive. Thus, the synchronization may not be successfully achieved. The loss of synchronism (LOS) is manifested as the detected frequency significantly deviating from the grid fundamental frequency, which can force the IBG to disconnect. In 2016 and 2017, 1,200 and 900 MW photovoltaic resource interruption events were caused by grid faults in California [2], [3]. It was reported that the disconnection partially resulted from the LOS [2], [3]. Such events had attracted considerable attention over the LOS issue.

To avoid loss of generation and provide ancillary support services, the IBG is required by mandatory grid codes to stay connected during grid faults and provide reactive current injection [4], [5]. In particular, asymmetrical fault ride-through capability and dual-sequence current injection requirements have been stipulated by the German latest grid codes [5]. With the benefit of the dual-sequence current injection, the positive-sequence grid voltage can be supported, and also the detrimental effects of the voltage imbalance can be reduced [6]. During the fault period, performing accurate reactive current injection necessities a successful synchronization. Meanwhile, the synchronization is also affected by the current injection as the IBG terminal voltage is subject to the voltage-current interaction on the grid impedance [7]–[9].

Under symmetrical grid faults, the positive-sequence synchronization is of concern. Most of the existing synchronization stability studies were accomplished under the assumption that the grid fault is symmetrical [7]–[21]. The modeling in [9]–[11] revealed that the motion equation governing the synchronization behavior of phase-locked loops (PLLs) and synchronous generators shows a similar form. Consequently, the use of conventional transient stability analysis methods (e.g., the equal-area criterion) in the IBG synchronization stability research was inspired. Following this inspiration, the stability mechanism and criterion [9]–[12] and the stabilization strategies or guidelines [9], [10], [12]–[14] have been reported. The equal-area criterion cannot take into account the impact of damping. To this end, the phase portrait approach was employed in [15] and [16] to analyze the impact of the damping of the PLL on the stability margin. In [17] and [18], the impact of the inner current control loop dynamics and the network dynamics on the synchronization stability was described and discussed. In [19], the impact of the outer control loop dynamics was evaluated by a stability margin index. Moreover, the synchronization behavior of multi-converter grid-tied systems was investigated in [20] and [21], where it was found that individual converters exhibit similar dynamic behaviors as they are all synchronized by PLLs with the grid. Toward symmetrical faults, the synchronization stability of IBG has been extensively studied.

This work was supported by the National Natural Science Foundation of China (U2066602, 52061635102). *(Corresponding author: Hua Geng.)*

X. He, C. He, and H. Geng are with the Department of Automation, Beijing National Research Center for Information Science and Technology, Tsinghua University, Beijing, 100084, China (e-mail: he-xq16@mails.tsinghua.edu.cn; hcj20@mails.tsinghua.edu.cn; genghua@tsinghua.edu.cn).

S. Pan is with the College of Electrical, Energy and Power Engineering, Yangzhou University, Yangzhou, 225000, China (e-mail: pss970503@icloud.com).

F. Liu is with the State Key Laboratory of Power Systems, Department of Electrical Engineering, Tsinghua University, Beijing, 100084, China (e-mail: lfeng@tsinghua.edu.cn).

Asymmetrical faults are more frequent than symmetrical ones [6]. Under asymmetrical faults, the negative-sequence voltage component appears. Both the positive- and negative-sequence synchronizations are of importance, as required by the dual-sequence current injection [22], [23]. Regarding the ability of IBG to remain in synchronism with the grid frequency in both the positive and negative sequences after being subjected to a disturbance, it is referred to as *dual-sequence synchronization stability* in this paper. It is known that the sequence networks are interconnected under asymmetrical fault conditions. Therefore, the positive- and negative-sequence components are coupled (namely the sequence coupling) [24]. Owing to the sequence coupling, the stability mechanism of the dual-sequence synchronization is much more complicated than that of the positive-sequence synchronization alone. For the conventional power system, the transient stability assessment (TSA) for asymmetrical faults is similar to that for symmetrical faults. This is because the positive-sequence synchronization alone is concerned. The impact of the negative- and zero-sequence network is taken into account by inserting an additional impedance into the positive-sequence network [25]. Collectively, very little is currently known about the dual-sequence synchronization stability of IBG. It remains unclear how to describe the dual-sequence synchronization behavior, what are the fundamental instability mechanism, and how to identify the stability conditions.

This paper aims to fill these gaps. By formulating the sequence voltage expression via the symmetrical components method, a dual-sequence synchronization model is developed. Considering that remaining in synchronism entails a feasible equilibrium point, the stability analysis mainly focuses on the existence of stable equilibrium points. By looking into the conditions for the equilibrium points, the types of instability due to the absence of equilibrium points are defined. Moreover, the impact mechanism of current references, circuit parameters, and fault types on different types of instability is investigated thoroughly, where the dominant instability factors are clarified and verified. The findings of this paper contribute to insightfully understanding the LOS event occurring when IBG rides through asymmetrical grid faults. The contributions provided in this paper can be summarized as follows:

- Presented a dual-sequence synchronization model that enables the quantitative analysis of stability;
- Identified the conditions for dual-sequence synchronization stability considering the sequence coupling;
- Defined and analyzed four instability types and quantified the dominant instability factors.

The rest of this paper proceeds as follows. In Section II, an overview of a dual-sequence current-controlled IBG system is presented and the grid code requirements are reviewed. The system is modeled in Section III. The stability conditions and instability types are identified in Section IV. Simulation and experimental verifications are conducted in Section V. The conclusion is drawn in Section VI.

## II. SYSTEM OVERVIEW AND GRID CODE REQUIREMENTS

A typical grid-connection topology and control diagram of an IBG system is illustrated in Fig. 1. Generally, the fundamental synchronization stability characteristics of actual multi-converter grid-connected systems can be understood with the knowledge gained from a single-converter system. Therefore, a single-converter prototype system is considered in this study. In the latest grid codes, e.g., German grid code VDE-AR-N 4120 [5], it is required that the generation installed in the high-voltage network can ride through asymmetrical faults and provide dynamic reactive current support in the positive and negative sequences. To this end, the IBG system in Fig. 1 should perform dual-sequence reactive current injection during asymmetrical grid faults. By doing this, the positive-sequence voltage can be boosted while the negative-sequence voltage can be reduced [26]. During the fault period, the reactive current references are directly specified according to the sequence voltage amplitude, as displayed in Fig. 2.

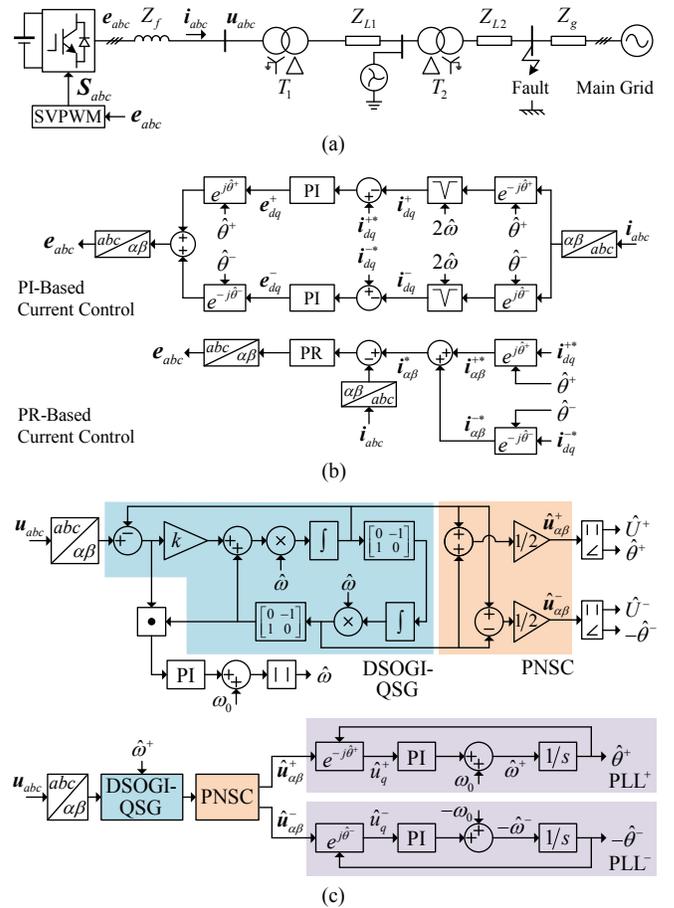

Fig. 1. Grid-connected IBG system under asymmetrical grid faults. (a) Circuit topology. (b) Proportional-integral (PI)-based or proportional-resonant (PR) regulator-based dual-sequence current control. (c) DSOGI-FLL/-PLL-based synchronization unit.

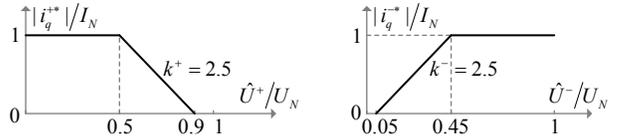

Fig. 2. Dual-sequence reactive current injection specifications based on the grid code [5].

Under asymmetrical fault conditions, a synchronization unit should detect both the positive- and negative-sequence voltage vector positions to conduct the dual-sequence current control simultaneously [22], [23]. Note that the negative-sequence control is dependent on the detection of the negative-sequence phase-angle, which cannot be replaced with the opposite of the positive-sequence phase-angle [22]. A commonly used synchronization unit is the dual second-order generalized integrator (DSOGI)-based phase-locked loop (DSOGI-PLL) [1] or frequency-locked loop (DSOGI-FLL) [27], which is depicted in Fig. 1(c). The DSOGI-PLL consists of the DSOGI quadrature signal generator (DSOGI-QSG, also called notch filter [27]) block, the positive- and negative-sequence calculation (PNSC) block, and the PLL block. The DSOGI-QSG block is utilized to generate an orthogonal version (90º-lagging) of the input signal. The sequence voltages can then be obtained with the decomposition by the PNSC block. The positive-/negative-sequence phase-angle is finally observed by the PLL$^+$/PLL$^-$ block, respectively. In the DSOGI-FLL, the phase-angle is calculated by algebraic manipulation instead. The tuning of the PLL/FLL parameters is of importance for its dynamic performance. The parameters can be carefully tuned according to the requirements of setting time, overshooting, harmonic suppression, etc. References [28], [29] provided details on how to tune the PLL/FLL parameters.

It should be noted that the synchronization stability of the IBG system is concerned with the ability to remain in frequency synchronization with the grid *during the fault period*. This is distinguished from the transient stability of conventional power systems, where the post-fault synchronizing ability is of main concern. With this in mind, the modeling and stability analysis in this study is concerning the *on-fault* period.

### III. System Modeling

In Fig. 1, the grid-connection transformer $T_2$ is generally of DYg type. Therefore, the high-voltage-side zero-sequence circuit is closed through the grounding connection of the neutral point. The low-voltage-side zero-sequence circuit is open so that the zero-sequence component is absent [6]. Considering

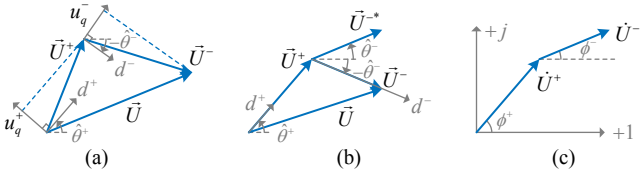

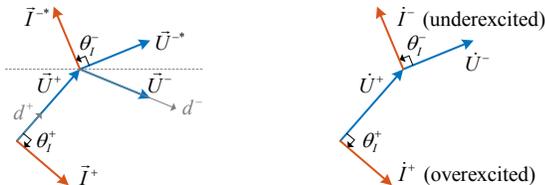

Fig. 4. Current vectors and phasors when injecting pure reactive current.

this fact, the IBG terminal phase voltages can be assumed to be

$$\begin{bmatrix} u_a \\ u_b \\ u_c \end{bmatrix} = \begin{bmatrix} U^+\cos(\omega t+\phi^+)+U^-\cos(\omega t+\phi^-) \\ U^+\cos(\omega t-2\pi/3+\phi^+)+U^-\cos(\omega t+2\pi/3+\phi^-) \\ U^+\cos(\omega t+2\pi/3+\phi^+)+U^-\cos(\omega t-2\pi/3+\phi^-) \end{bmatrix}. \quad (1)$$

The sequence voltage vectors (complex vectors) are denoted by $\vec{U}^+ = U^+ e^{j(\omega t+\phi^+)}$, $\vec{U}^- = U^- e^{-j(\omega t+\phi^-)}$. The sequence voltage phasors are defined via the symmetrical components method as $\dot{U}^+ = U^+ e^{j\phi^+}$, $\dot{U}^- = U^- e^{j\phi^-}$. The relationship between the vectors and the phasors are then expressed by [23],

$$\vec{U}^+ = \dot{U}^+ e^{j\omega t}, \vec{U}^{-*} = \dot{U}^- e^{j\omega t}. \quad (2)$$

The complex conjugate operation "*" is applied to the neg-ative-sequence voltage vector as it rotates clockwise with time, as shown in Fig. 3. The positive- and negative-sequence phase-angle positions estimated by the synchronization unit are represented by $d^+$ and $d^-$. In a steady state, $d^+$ and $d^-$ indicate the direction of the corresponding voltage vectors. The steady state further gives rise to the phasor diagram, as seen in Fig. 3(c). While in the dynamic response, there are some deviations in the estimated frequency and phase-angle, as seen in Fig. 3(a). Additionally, the IBG terminal voltage changes with the output current during dynamic conditions. Therefore, the model describing the IBG synchronization behavior is composed of the terminal voltage representation and the synchronization unit.

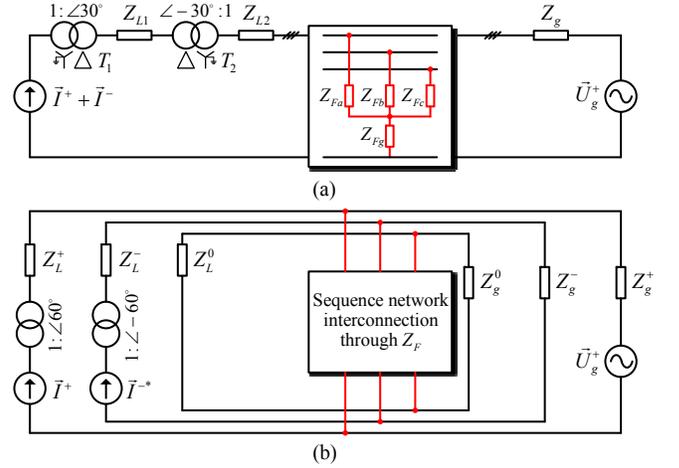

Fig. 5. (a) Phase-domain circuit. (b) Sequence-domain circuit, where $Z_L^\pm$ is contributed by $Z_{L1}$, $Z_{L2}$ and the two transformers; $Z_L^0$ is only contributed by $Z_{L2}$ and $T_2$. Considering the complete transposition of transmission lines, the positive- and negative-sequence line impedances are equal, denoted by $Z_g^+ = Z_g^- = |Z_g| \angle \phi_g$, $Z_L^+ = Z_L^- = |Z_L| \angle \phi_L$. Subfigure (b) is detailed in Fig. 6.

TABLE I
IMPEDANCE OF THE FAULT BRANCH

| Fault type | $Z_{Fa}$ | $Z_{Fb}$ | $Z_{Fc}$ | $Z_{Fg}$ |
|---|---|---|---|---|
| SLG | 0 | ∞ | ∞ | $Z_F$ |
| DLG | ∞ | 0 | 0 | $Z_F$ |
| LL | ∞ | $Z_F/2$ | $Z_F/2$ | ∞ |

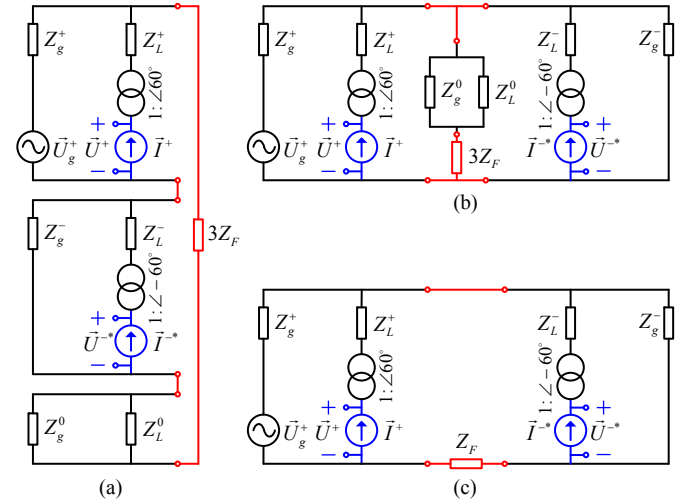

Fig. 6. Sequence-domain circuit under different types of faults. (a) SLG-type fault. (b) DLG-type fault. (c) LL-type fault.

### A. Terminal Voltage Representation

When investigating large-disturbance synchronization stability, the PLL/FLL dynamics are of special concern, whereas the current control dynamics and the network transients can be ignored to simplify the modeling [9]–[13]. This simplification is justified by the fact that the synchronization dynamics are involved with the bandwidth of the synchronization unit, which is generally ten times smaller than the current controller bandwidth. Therefore, the synchronization dynamics can be approximately decoupled from the current dynamics and analyzed separately in large-disturbance stability studies [9]–[13].

Without considering the current control dynamics, a quasi-steady-state current source can be used to represent the inverter. The orientation for the output current is provided by the synchronization unit. The angles are denoted by $\hat{\theta}^+$ and $-\hat{\theta}^-$. The relative angles of the current vectors to the rotating reference frames are denoted by $\theta_I^+$ and $-\theta_I^-$. Accordingly, the positive- and negative-sequence current outputs are defined by,

$$\vec{I}^+ = I^+ e^{j(\hat{\theta}^+ + \theta_I^+)} \qquad (3)$$
$$\vec{I}^- = I^- e^{-j(\hat{\theta}^- + \theta_I^-)}.$$

If the active current is zero, the positive-sequence reactive current output is overexcited, and the negative-sequence one is underexcited, then $\theta_I^+ = -\pi/2$, $\theta_I^- = \pi/2$, as seen in Fig. 4.

The phase-domain and sequence-domain circuits under any type of fault are shown in Fig. 5. Note that in addition to the winding configuration, the transformer model is also related to the composition and the core structure. A transformer composed of three single-phase transformers is considered. The transformer phase shift is included in the circuit as it has an opposite effect on the positive- and negative-sequence voltages. Since the per-unit presentation is used, the transformation ratio is eliminated. It is suggested by (2) that the negative-sequence vector in Fig. 5(b) should be expressed in the form of complex conjugate. Also, it is seen that the occurrence of asymmetrical faults causes the coupling between the sequence components.

The fault impedance is summarized in Table I in the case of single line-to-ground (SLG), double line-to-ground (DLG), and line-to-line (LL) faults. Without loss of generality, phase A is considered a special phase in the faults. $Z_F$ is used to represent the fault impedance. Under different types of faults, the boundary conditions at the fault node can be formulated using the symmetrical components method [30]. Then, the sequence-domain circuit can be drawn, as seen in Fig. 6. Note that the presented circuit here distinguishes itself from the conventional established one in the asymmetrical network analysis [30]. The major difference is that the dual-sequence current injection is incorporated into the circuit representation.

Based on the sequence-domain circuit, the inverter terminal sequence voltage can be represented. If it is assumed that the current reference remains constant during grid faults, $I^+ e^{j\theta_I^+}$ and $I^- e^{j\theta_I^-}$ will remain constant as well because the current control dynamics are ignored. Under this assumption, the voltage drop produced by $\vec{I}^+$ on a branch with resistance $R$ and inductance $L$ is $\Delta \vec{U}_Z^+ = R\vec{I}^+ + L d\vec{I}^+/dt = R\vec{I}^+ + j\hat{\omega}L\vec{I}^+ = Z\vec{I}^+$ [9], where $Z = R + j\hat{\omega}^+ L$, $\hat{\omega}^+ = d\hat{\theta}^+/dt$. Similarly, the voltage drop produced by $\vec{I}^-$ is $\Delta \vec{U}_Z^{-*} = Z\vec{I}^{-*}$. In this way, the voltage equation of the circuit can be formed, as given by,

$$\vec{U}^+ e^{j\pi/3} = K_1 \vec{U}_g^+ + Z_2 \vec{I}^+ e^{j\pi/3} + Z_3 \vec{I}^{-*} e^{-j\pi/3} \qquad (4)$$

$$\vec{U}^{-*} e^{-j\pi/3} = K_4 \vec{U}_g^+ + Z_5 \vec{I}^{-*} e^{-j\pi/3} + Z_6 \vec{I}^+ e^{j\pi/3} \qquad (5)$$

where $K_i = |K_i|\angle\phi_i$, $i = 1, 4$ and $Z_i = |Z_i|\angle\phi_i$, $i = 2, 3, 5, 6$ are summarized in Table II. In (4), (5), both the voltage components $K_1 \vec{U}_g^+$ and $K_4 \vec{U}_g^+$ are produced from the grid voltage $\vec{U}_g^+$. The latter is derived from the sequence network interconnection. Besides this, the sequence coupling is also reflected by the terms relevant to $Z_3$ and $Z_6$. This coupling stems from the dual-sequence current injection and results in the interaction between the voltage and current components. Equations (4) and (5) further lead to (6) and (7),

$$\vec{U}^+ = |K_1|U_g^+ \angle(\phi_1 + \theta_g^+ - \pi/3) + |Z_2|I^+ \angle(\phi_2 + \hat{\theta}^+ + \theta_I^+)$$
$$+ |Z_3|I^- \angle(\phi_3 + \hat{\theta}^- + \theta_I^- - 2\pi/3) \quad (6)$$

$$\vec{U}^{-*} = |K_4|U_g^+ \angle(\phi_4 + \theta_g^+ + \pi/3) + |Z_5|I^- \angle(\phi_5 + \hat{\theta}^- + \theta_I^-)$$
$$+ |Z_6|I^+ \angle(\phi_6 + \hat{\theta}^+ + \theta_I^+ + 2\pi/3) \quad (7)$$

TABLE II
EXPRESSIONS OF $|K_i|\angle\phi_i$ AND $|Z_i|\angle\phi_i$ UNDER DIFFERENT TYPES OF FAULTS

| Symbol | SLG | DLG | LL | 3LG |
|---|---|---|---|---|
| $K_1 = \|K_1\|\angle\phi_1$ | $\dfrac{Z_g^- + Z_g^0 \mathbin{/\mkern-6mu/} Z_L^0 + 3Z_F}{Z_g^+ + Z_g^- + Z_g^0 \mathbin{/\mkern-6mu/} Z_L^0 + 3Z_F}$ | $\dfrac{Z_g^0 \mathbin{/\mkern-6mu/} Z_L^0 + 3Z_F}{Z_g^+ + 2(Z_g^0 \mathbin{/\mkern-6mu/} Z_L^0) + 6Z_F}$ | $\dfrac{Z_g^- + Z_F}{Z_g^+ + Z_g^- + Z_F}$ | $\dfrac{Z_F}{Z_g^+ + Z_F}$ |
| $Z_2 = \|Z_2\|\angle\phi_2$ | $\dfrac{Z_g^+(Z_g^- + Z_g^0 \mathbin{/\mkern-6mu/} Z_L^0 + 3Z_F)}{Z_g^+ + Z_g^- + Z_g^0 \mathbin{/\mkern-6mu/} Z_L^0 + 3Z_F} + Z_L^+$ | $\dfrac{Z_g^+(Z_g^0 \mathbin{/\mkern-6mu/} Z_L^0 + 3Z_F)}{Z_g^+ + 2(Z_g^0 \mathbin{/\mkern-6mu/} Z_L^0) + 6Z_F} + Z_L^+$ | $\dfrac{Z_g^+(Z_g^- + Z_F)}{Z_g^+ + Z_g^- + Z_F} + Z_L^+$ | $\dfrac{Z_g^+ Z_F}{Z_g^+ + Z_F} + Z_L^+$ |
| $Z_3 = \|Z_3\|\angle\phi_3$ | $-\dfrac{Z_g^+ Z_g^-}{Z_g^+ + Z_g^- + Z_g^0 \mathbin{/\mkern-6mu/} Z_L^0 + 3Z_F}$ | $\dfrac{Z_g^+(Z_g^0 \mathbin{/\mkern-6mu/} Z_L^0 + 3Z_F)}{Z_g^+ + 2(Z_g^0 \mathbin{/\mkern-6mu/} Z_L^0) + 6Z_F}$ | $\dfrac{Z_g^+ Z_g^-}{Z_g^+ + Z_g^- + Z_F}$ | 0 |
| $K_4 = \|K_4\|\angle\phi_4$ | $-\dfrac{Z_g^-}{Z_g^+ + Z_g^- + Z_g^0 \mathbin{/\mkern-6mu/} Z_L^0 + 3Z_F}$ | $\dfrac{Z_g^0 \mathbin{/\mkern-6mu/} Z_L^0 + 3Z_F}{Z_g^+ + 2(Z_g^0 \mathbin{/\mkern-6mu/} Z_L^0) + 6Z_F}$ | $\dfrac{Z_g^-}{Z_g^+ + Z_g^- + Z_F}$ | 0 |
| $Z_5 = \|Z_5\|\angle\phi_5$ | $\dfrac{Z_g^-(Z_g^+ + Z_g^0 \mathbin{/\mkern-6mu/} Z_L^0 + 3Z_F)}{Z_g^+ + Z_g^- + Z_g^0 \mathbin{/\mkern-6mu/} Z_L^0 + 3Z_F} + Z_L^-$ | $\dfrac{Z_g^-(Z_g^0 \mathbin{/\mkern-6mu/} Z_L^0 + 3Z_F)}{Z_g^- + 2(Z_g^0 \mathbin{/\mkern-6mu/} Z_L^0) + 6Z_F} + Z_L^-$ | $\dfrac{Z_g^-(Z_g^+ + Z_F)}{Z_g^+ + Z_g^- + Z_F} + Z_L^-$ | 0 |
| $Z_6 = \|Z_6\|\angle\phi_6$ | $-\dfrac{Z_g^+ Z_g^-}{Z_g^+ + Z_g^- + Z_g^0 \mathbin{/\mkern-6mu/} Z_L^0 + 3Z_F}$ | $\dfrac{Z_g^+(Z_g^0 \mathbin{/\mkern-6mu/} Z_L^0 + 3Z_F)}{Z_g^+ + 2(Z_g^0 \mathbin{/\mkern-6mu/} Z_L^0) + 6Z_F}$ | $\dfrac{Z_g^+ Z_g^-}{Z_g^+ + Z_g^- + Z_F}$ | 0 |

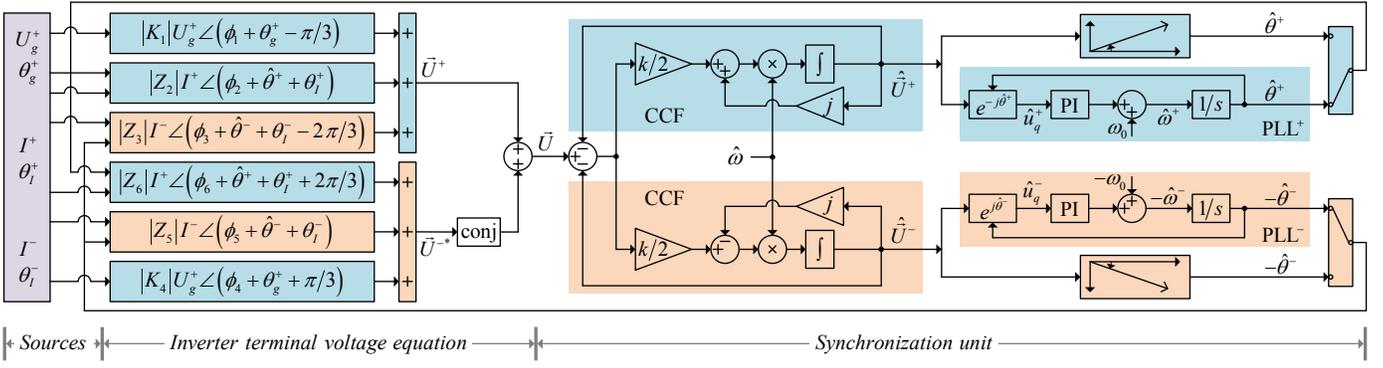

Fig. 7. Proposed model for quantitatively analyzing dual-sequence synchronization stability of the IBG system, where the complex-coefficient filter (CCF) is used in the synchronization unit in order to intuitively show the sequence coupling. Note that there are two options in the estimation of the angles (by arctangent functions or PLL$^{+/-}$), which lead to different dynamic behaviors.

The IBG terminal voltage is finally represented by,

$$\vec{U} = \vec{U}^+ + \text{conj}(\vec{U}^{-*}) \quad (8)$$

### B. Synchronization Unit Modeling

The synchronization unit modeling has been well documented in the literature [1], [27]–[29]. The DSOGI-QSG block can be expressed in the form of complex vectors by,

$$\begin{cases} d\hat{\vec{U}}/dt = j\hat{\omega}\hat{\vec{V}} + k\hat{\omega}(\vec{U} - \hat{\vec{U}}) \\ d\hat{\vec{V}}/dt = j\hat{\omega}\hat{\vec{U}} \end{cases} \quad (9)$$

where $\hat{\vec{U}}$ and $\hat{\vec{V}}$ denote the filtered version and the 90° lagging version of the input signal $\vec{U}$, respectively. A frequency adaptive loop can be employed in the DSOGI-QSG block to adapt to the grid frequency variation, which is defined by [27],

$$\hat{\omega} = k_p \, \text{Im}\left[(\vec{U} - \hat{\vec{U}})\hat{\vec{V}}^*\right] + k_i \int_0^t \text{Im}\left[(\vec{U} - \hat{\vec{U}})\hat{\vec{V}}^*\right] d\tau + \omega_0. \quad (10)$$

The PNSC block is given by,

$$\hat{\vec{U}}^+ = 1/2(\hat{\vec{U}} + \hat{\vec{V}}), \quad \hat{\vec{U}}^- = 1/2(\hat{\vec{U}} - \hat{\vec{V}}). \quad (11)$$

The DSOGI-QSG is a real-coefficient filter (RCF) [27]. The complex-coefficient filter (CCF)-based synchronization algorithm has also been well documented in the literature. According to the report in [27], the RCF-based FLL/PLL and CCF-based FLL/PLL are mathematically equivalent. The demonstration is a straightforward process by considering (11) as a linear transformation to (9). For ease of comparative analysis, the equation of the CCF is given as follows.

$$\begin{cases} d\hat{\vec{U}}^+/dt = j\hat{\omega}\hat{\vec{U}}^+ + k\hat{\omega}/2(\vec{U} - \hat{\vec{U}}^+ - \hat{\vec{U}}^-) \\ d\hat{\vec{U}}^-/dt = -j\hat{\omega}\hat{\vec{U}}^- + k\hat{\omega}/2(\vec{U} - \hat{\vec{U}}^+ - \hat{\vec{U}}^-) \end{cases} \quad (12)$$

After the positive- and negative-sequence components are estimated, the angles can be either computed by an arctangent function or identified by the PLL$^{+/-}$ block. The PLL$^+$ and PLL$^-$ blocks are expressed by (13) and (14), respectively,

$$\begin{cases} d\hat{\theta}^+/dt = \hat{\omega}^+ \\ d\hat{\omega}^+/dt = k_p \hat{u}_q^+ + k_i \int \hat{u}_q^+ dt + \omega_0 \end{cases} \quad (13)$$

$$\begin{cases} d\hat{\theta}^-/dt = \hat{\omega}^- \\ d\hat{\omega}^-/dt = -k_p \hat{u}_q^- - k_i \int \hat{u}_q^- dt + \omega_0 \end{cases} \quad (14)$$

where $\hat{u}_q^+$ and $\hat{u}_q^-$ are q-axis voltages, as seen in Fig. 3(a).

### C. Dual-Sequence Synchronization Model of the IBG System

By connecting the inverter terminal voltage equation with the synchronization unit model, the dual-sequence synchronization model is developed, as shown in Fig. 7. The model paves the way for quantitatively investigating the synchronization behavior of a dual-sequence current-controlled IBG system. It is noted that the model is nonlinear and hence applies to large-disturbance stability analysis. Also, it is general to asymmetrical and symmetrical short-circuit fault types. In this sense, the previously established model for symmetrical grid faults [9] is a special case of this general model. Note that since a current source is used to represent the IBG and the current reference can be specified directly, the current limiter is ignored in the model. Without the current limiter, the maximum allowable current injection constrained by the synchronization stability can be fully analyzed and verified, which can be seen in the next section.

The accuracy of the model is verified by comparison with a detailed simulation model. The simulation model is represented by an average-value model of inverters, in which the PLL/FLL dynamics, current control dynamics, and network dynamics are included. The simulation is accomplished with a fixed 2e–5 seconds step size and a discrete solver. The results are shown in Fig. 8. It is seen that the proposed model is able to approximate the detailed model quite well. In the on-fault steady state, in particular, the results of the two models are highly coincident, which verifies that the voltage equations derived in (6) and (7) are correct. There are deviations at the moment of the fault occurrence and clearance. This is mainly because the current control dynamics and the network transients are ignored, as discussed in [17] and [18].

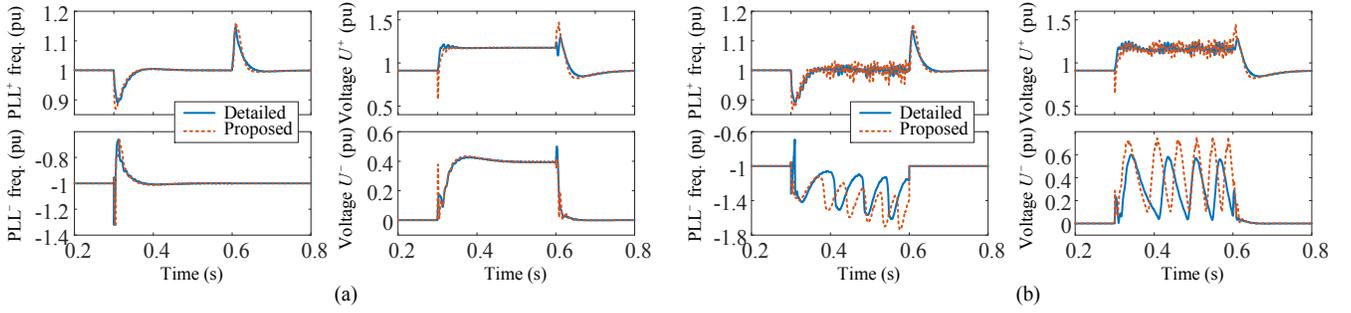

Fig. 8. Verification results of the proposed model under the SLG fault. (a) Stable case. (b) Unstable case.

It is worth reminding that a dual-sequence synchronization model was reported in [31] recently. However, it was shown in [31] that the model exhibits some steady-state error. This was in fact due to the improper use of the sequence network split to derive the Thévenin equivalents. As a consequence, the sequence network interconnection was broken and the circuit operating point was also altered in the derivation. The steady-state error is addressed in the model developed here by avoiding using the sequence network split.

For asymmetrical grid faults, the multi-converter synchronization model can be established similarly. Referring to the single-converter model in Fig. 7, a few expected remarks can be drawn as follows. Firstly, the synchronization of one converter is impacted by the output current of the others due to the grid impedance interaction. Secondly, the converters have similar synchronization behaviors as the same grid-following mechanism is adopted. The multi-converter synchronization model under symmetrical faults has been reported in [20], where the research results are consistent with these expected remarks. More insights into the multi-converter dual-sequence synchronization will be provided in future work.

## IV. Synchronization Stability Analysis

The IBG is supposed to remain in synchronism with the grid during the fault period. To achieve this, it should be ensured at first that there is a feasible steady-state operating point. On this basis, nonlinear stability analysis can be applied to estimate the domain of attraction (also referred to as stability region) [11], [14]; eigenvalue analysis, impedance analysis, etc. [32], [33] can be applied to examine the small-signal stability margin. In this study, the stability analysis mainly focuses on the characteristics of equilibrium points, providing a steady-state basis for nonlinear and small-signal stability studies.

### A. Conditions for Equilibrium Points

Using $e^{-j\hat{\theta}^+}$, $e^{-j\hat{\theta}^-}$ to transform (6), (7) into the rotating reference frames leads to (15), (16), respectively,

$$u_d^+ + ju_q^+ = |K_1||U_g^+|\angle(\phi_1 - \delta^+) + |Z_2||I^+|\angle(\phi_2 + \theta_I^+) \\ + |Z_3||I^-|\angle(\phi_3 + \theta_I^- - \delta^+ + \delta^-) \quad (15)$$

$$u_d^- - ju_q^- = |K_4||U_g^+|\angle(\phi_4 - \delta^-) + |Z_5||I^-|\angle(\phi_5 + \theta_I^-) \\ + |Z_6||I^+|\angle(\phi_6 + \theta_I^+ + \delta^+ - \delta^-) \quad (16)$$

where $\delta^+ = \hat{\theta}^+ - \theta_g^+ + \pi/3$, $\delta^- = \hat{\theta}^- - \theta_g^+ - \pi/3$. In a steady state (if it exists), $u_q^\pm = 0$ and $u_d^\pm > 0$ are the basic requirements for the voltage vector orientation, i.e.,

$$\begin{cases} |K_1||U_g^+|\sin(\phi_1 - \delta^+) + |Z_2||I^+|\sin(\phi_2 + \theta_I^+) + \\ \quad |Z_3||I^-|\sin(\phi_3 + \theta_I^- - \delta^+ + \delta^-) = 0 \\ |K_4||U_g^+|\sin(\phi_4 - \delta^-) + |Z_5||I^-|\sin(\phi_5 + \theta_I^-) + \\ \quad |Z_6||I^+|\sin(\phi_6 + \theta_I^+ + \delta^+ - \delta^-) = 0. \end{cases} \quad (17)$$

$$\begin{cases} |K_1||U_g^+|\cos(\phi_1 - \delta^+) + |Z_2||I^+|\cos(\phi_2 + \theta_I^+) + \\ \quad |Z_3||I^-|\cos(\phi_3 + \theta_I^- - \delta^+ + \delta^-) > 0 \\ |K_4||U_g^+|\cos(\phi_4 - \delta^-) + |Z_5||I^-|\cos(\phi_5 + \theta_I^-) + \\ \quad |Z_6||I^+|\cos(\phi_6 + \theta_I^+ + \delta^+ - \delta^-) > 0 \end{cases} \quad (18)$$

The physical meaning of (17) and (18) is interpreted as follows. The IBG is represented with a current source. The terminal voltage and the output current are subject to Kirchhoff's laws while the output current control is bound by the voltage orientation. If the solution for $\delta^+$ and $\delta^-$ is nonexistent, the dual-sequence synchronization along with the current control cannot be fulfilled. This will consequently result in the LOS. Therefore, the nonexistence of steady-state equilibrium points is one of the underlying causes of synchronization instability.

Generally, as long as there are solutions to meet (17) and (18), the solutions can be classified into stable equilibrium points and unstable equilibrium points according to the symmetry of trigonometric function [12]. In other words, if there are equilibrium points, there should be a stable equilibrium point. The point can be identified with the negative feedback condition of the synchronization unit, $du_q^+/d\delta^+ < 0$, $-du_q^-/d\delta^- < 0$, i.e.,

$$\begin{cases} -|K_1||U_g^+|\cos(\phi_1 - \delta^+) - |Z_3||I^-|\cos(\phi_3 + \theta_I^- - \delta^+ + \delta^-) < 0 \\ -|K_4||U_g^+|\cos(\phi_4 - \delta^-) - |Z_6||I^+|\cos(\phi_6 + \theta_I^+ + \delta^+ - \delta^-) < 0. \end{cases} \quad (19)$$

The conditions that stable equilibrium points should follow are proposed by (17) to (19). By numerically solving (17) under the constraints of (18) and (19), the on-fault steady-state point can be calculated. To facilitate an intuitive understanding of the steady-state point, the phasor diagram is drawn, as depicted in Fig. 9, where the circuit parameters are provided in Table III. As expected, the positive-sequence terminal voltage is boosted with an overexcited current injection; the negative-sequence one is reduced with an underexcited current injection. Also, it is important to note that the sequence coupling is counterproductive to the voltage boost and reduction. This is manifested as $|Z_3||I^-|\angle(\phi_3 + \theta_I^- + \delta^-)$ reacting against $|K_1||U_g^+|\angle\phi_1$ whereas $|Z_6||I^+|\angle(\phi_6 + \theta_I^+ + \delta^+)$ enhancing support for $|K_4||U_g^+|\angle\phi_4$.

As discussed by the previous research [34], the absence of the equilibrium point implies that the closed polygon in Fig. 9 cannot be formed by the phasor synthesis. This can lead to an inappropriate interaction appearing between the controllers and the grid impedance [34]. The interaction cannot settle down due to the absence of equilibrium points, thus followed by a synchronization instability phenomenon. It should be noted that as long as one of the conditions in (17) to (19) cannot be followed, both the dual-sequence current control will lose synchronism. This is due to the sequence coupling effect. From the phasor diagram in Fig. 9, it can be found that the instability is related to fault types, current references, and circuit parameters. A comprehensive investigation on the possible instability types and dominant factors is launched below.

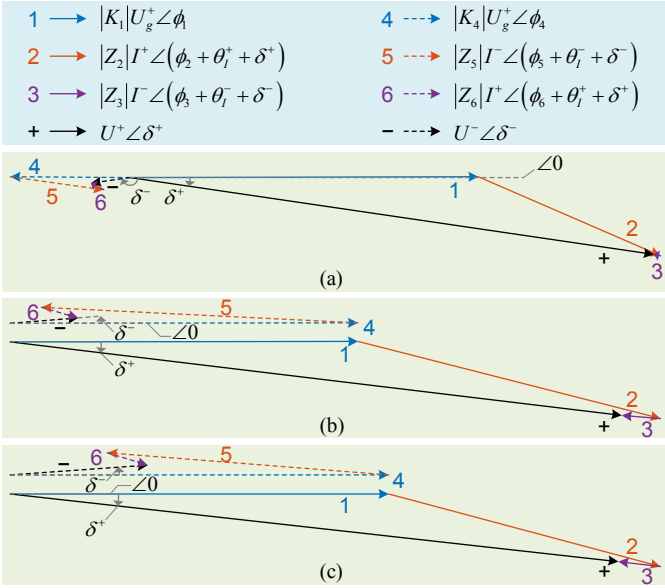

Fig. 9. Phasor diagrams of the on-fault steady state. (a) SLG fault. (b) DLG fault. (c) LL fault. In the SLG fault, $I^+\angle\theta_I^+ = 0.6\angle -90°$, $I^-\angle\theta_I^- = 0.3\angle 90°$. In the DLG and LL fault, $I^+\angle\theta_I^+ = 0.6\angle -90°$, $I^-\angle\theta_I^- = 0.6\angle 90°$. Note that $\delta^+$ and $\delta^-$ in the phasor diagrams denote the angle of $U^+\angle\delta^+$ and $U^-\angle\delta^-$ relative to the zero-degree direction, respectively.

TABLE III
SYSTEM PARAMETERS

| Symbol | Description | Value |
|---|---|---|
| $U_N$ | Voltage level | 110 kV / 35 kV / 690 V |
| $U_g$ | Grid voltage | 120 kV |
| $S_N$ | Nominal capacity | 9 MVA |
| $f_n$ | Nominal frequency | 50 Hz |
| $Z_f$ | Choke impedance | 0.15/50 + j0.15 p.u. |
| $Z_{T1}^+ = Z_{T1}^- = Z_{T1}^0$ | $T_1$ impedance | 0.06/30 + j0.06 p.u. |
| $Z_{T2}^+ = Z_{T2}^- = Z_{T2}^0$ | $T_2$ impedance | 0.16/30 + j0.16 p.u. |
| $Z_{L1}^+ = Z_{L1}^- = Z_{L1}^0/3$ | $Z_{L1}$ impedance | 0.02 + j0.05 p.u. |
| $Z_{L2}^+ = Z_{L2}^- = Z_{L2}^0/3$ | $Z_{L2}$ impedance | 0.06 + j0.3 p.u. |
| $Z_g^+ = Z_g^- = Z_g^0/3$ | $Z_g$ impedance | 0.04 + j0.2 p.u. |
| $Z_F$ | Fault impedance | 0.01 Ω |
| $k$ | Gain of SOGI unit | 1.414 |
| $K_{p\_PLL}, K_{i\_PLL}$ | PLL PI parameters | 100, 2000 |
| $K_{p\_Curr}, K_{r\_Curr}, \omega_C$ | PR parameters | 10, 200, 10 |
| $f_{sw}$ | Switching frequency | 4 kHz |
| $T_s$ | Sampling period | 1e–4 s |

### B. Instability Types and Dominant Factors

Owing to the complexity of trigonometric function, figuring out the explicit conditions for one specific parameter is a non-trivial task. For this reason, the dominant instability factors will be ascertained by a decoupling analysis at first. Then the impact of the sequence coupling on the stability will be investigated.

In Table II, $Z_2$ and $Z_5$ contain the grid-connection impedance $|Z_L|$. The size of $|Z_2|$ and $|Z_5|$ is therefore much larger than that of $|Z_3|$ and $|Z_6|$, especially in a weak connection scenario, as indicated by the phasor diagram in Fig. 9. Ignoring the sequence coupling terms involving $Z_3$ and $Z_6$ in (17), the conditions for the existence of equilibrium points can be formulated explicitly, as given by,

$$|Z_2|I^+ \sin|\phi_2 + \theta_I^+| \leq |K_1||U_g^+| \Rightarrow I_{\lim}^+ = \frac{|K_1||U_g^+|}{|Z_2|\sin|\phi_2 + \theta_I^+|} \quad (20a)$$

$$|Z_5|I^- \sin|\phi_5 + \theta_I^-| \leq |K_4||U_g^+| \Rightarrow I_{\lim}^- = \frac{|K_4||U_g^+|}{|Z_5|\sin|\phi_5 + \theta_I^-|} \quad (20b)$$

where the boundary conditions are expressed by the limit on the current injection amplitude. Analogously, ignoring the sequence coupling terms in (18) and then combining with (17) can formulate the other part of the boundary conditions,

$$|Z_2|I_{\lim}^+ = |K_1||U_g^+| \Rightarrow I_{\lim}^+ = |K_1||U_g^+|/|Z_2| \quad (21a)$$

$$|Z_5|I_{\lim}^- = |K_4||U_g^+| \Rightarrow I_{\lim}^- = |K_4||U_g^+|/|Z_5| \quad (21b)$$

The boundary conditions are exemplified in Figs. 10(a) and 11(a). The shadow area represents the allowable region for the current injection. The solid-line boundaries of Regions 1 and 3 are defined by (20a) and (20b), respectively. The dashed-line boundaries of Regions 2 and 4 are defined by (21a) and (21b), respectively. Observing these boundaries, the instability types and dominant factors can be summarized as follows.

*1) Positive Sequence Dominated Type-1 Instability:* i.e., the instability due to the positive-sequence current injection going beyond the boundary of Region 1. The current injection is basically of over-excitation in this case. As seen in Fig. 10(b), assuming the voltage vector $\vec{U}^+$ rotates to the zero-degree angle at a certain moment, the impedance voltage drop $Z_2\vec{I}_{\lim}^+$ can be drawn under the current injection $\vec{I}_{\lim}^+$. Centered on the endpoint of $Z_2\vec{I}_{\lim}^+$, a circle with a radius equal to $|K_1||U_g^+|$ is drawn. At the critical operating point, the circle is tangent to the zero-degree horizontal line. If the current amplitude increases to exceed the limit, the horizontal line will fail to intersect with the circle and consequently, there will be no steady-state operating points. For this type of instability, the dominant factors are reflected by the current injection limit in (20a). In particular, the smaller the ratio $|K_1||U_g^+|/|Z_2|$ is, the smaller the allowable operating region as well. This suggests that both a severe grid fault and a weak grid connection are detrimental to synchronization stability.

*2) Positive Sequence Dominated Type-2 Instability:* i.e., the instability due to the positive-sequence current injection going beyond the boundary of Region 2. The current injection is mostly of under-excitation in this case. As depicted in Fig. 10(c), the voltage component $K_1\vec{U}_g^+ e^{-j\pi/3}$ is counterbalanced by the impedance voltage drop, i.e., the terminal voltage is reduced to zero. The current injection exceeding the limit is disallowed as it leads to a reverse (180-degree) terminal voltage vector. This will mislead and destabilize the voltage-oriented current control. The dominant factors of this instability type are reflected by the current injection limit in (21a). In contrast to the type-1 instability, $\phi_2 + \theta^+_I$ does not affect the limit here.

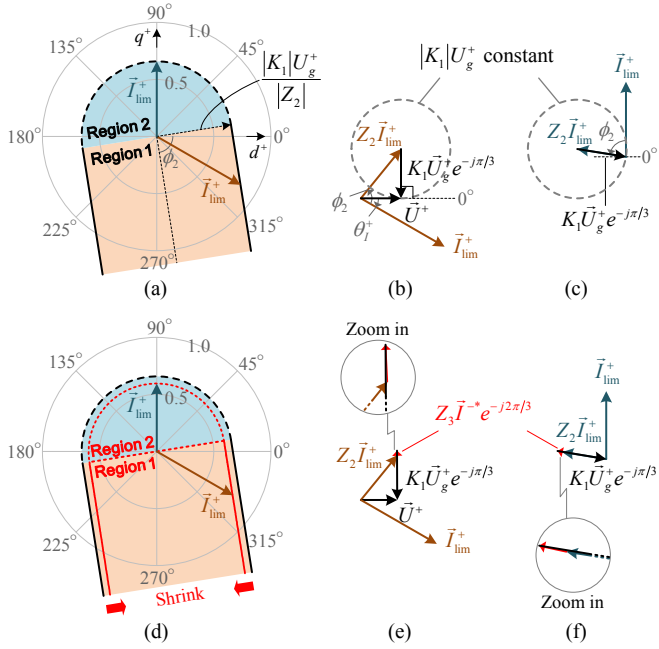

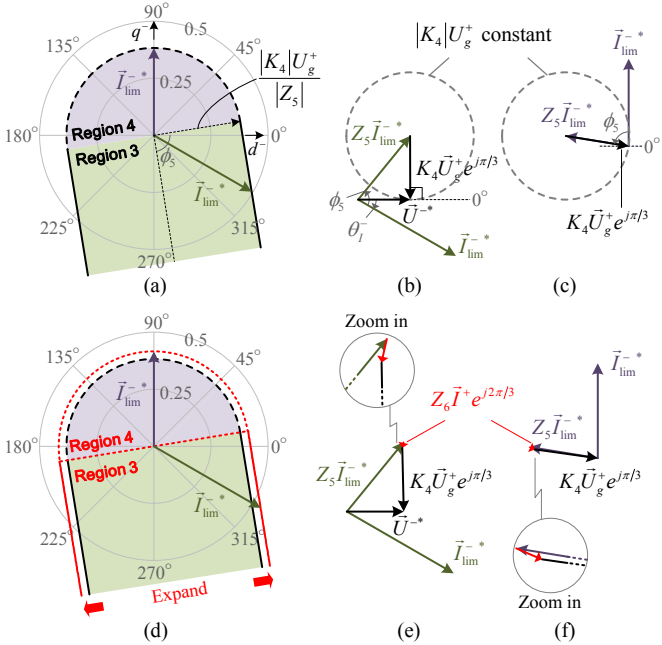

Fig. 10. (a) The allowable region for the positive-sequence current injection during a DLG-type fault. (b) For $\vec{I}^+_{lim}$ in Region 1, $|K_1|U^+_g$ matches just enough $Z_2\vec{I}^+_{lim}$ in the vertical direction. (c) For $\vec{I}^+_{lim}$ in Region 2, the terminal voltage is reduced to zero just right. (d) The allowable region for the positive-sequence current injection shrinks when considering the impact of a negative-sequence current injection, $I^-\angle\theta^-_I = 0.5\angle 90°$. Subfigures (e) and (f) show the vector diagrams depicting the equilibrium points on the boundary.

Fig. 11. (a) The allowable region for the negative-sequence current injection during a SLG-type fault. In (b) and (c), the critical operating points are explained similarly to those in Fig. 10. (d) The allowable region for the negative-sequence current injection expands when considering the impact of a positive-sequence current injection, $I^+\angle\theta^+_I = 0.5\angle -90°$. Subfigures (e) and (f) show the vector diagrams depicting the equilibrium points on the boundary.

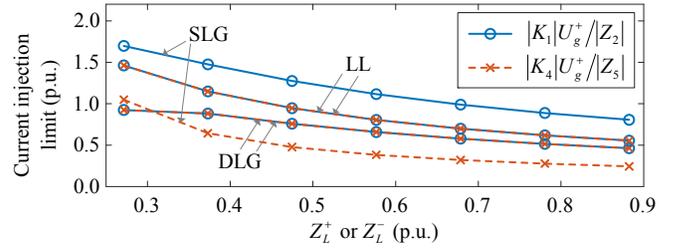

Fig. 12. Comparisons of the current injection limit under different fault types and different line impedances.

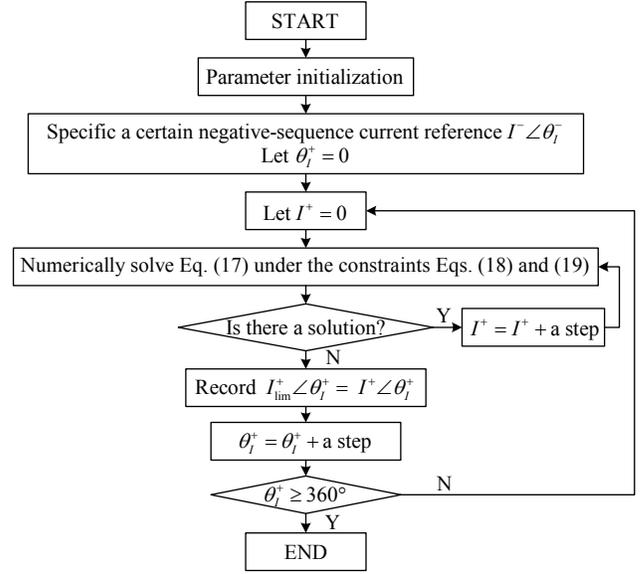

Fig. 13. Proposed traversal algorithm to identify the current injection limit considering the sequence coupling, where the identification of the positive-sequence current injection limit is considered as an example.

*3) Negative Sequence Dominated Type-1 Instability:* i.e., the instability due to the negative-sequence current injection going beyond the boundary of Region 3. The current injection is mainly of over-excitation. The critical operating point is demonstrated in Fig. 11(b). The dominant instability factors are reflected by the current injection limit in (20b), where the ratio $|K_4|U^+_g/(|Z_5|\sin|\phi_5 + \theta^-_I|)$ plays an important role in affecting the existence of equilibrium points. Due to $|K_4| \leq |K_1|$ whereas $|Z_2| = |Z_5|$, this type of instability is more likely than the positive-sequence dominated type-1 instability. This is the case under the SLG-type fault, as demonstrated by the results in Fig. 12.

*4) Negative Sequence Dominated Type-2 Instability:* i.e., the instability due to the positive-sequence current injection going beyond the boundary of Region 4. The current injection is mainly of under-excitation. Similarly, the boundary is shaped by the critical operating points where the negative-sequence voltage is reduced to zero. The ratio $|K_4|U^+_g/|Z_5|$ proves to be the dominant instability factor.

Four types of instability and the dominant factors have been clarified, which give insights into the impact mechanism of the current injection and the circuit impedances on the steady-state allowable operating region. In some of the present grid codes, it is important to note that the current injection is specified in the positive sequence only. This is actually a special case of the above analysis as only the positive-sequence dominated type-1 and type-2 instability is involved. Note also that the operating range exemplified in Figs. 10 and 11 is for the DLG and SLG fault types, respectively. The comparisons between different fault types are displayed in Fig. 12. As reflected by the ratio, the positive-sequence dominated instability is more likely to occur under the DLG-type fault compared to the other two types of fault. Furthermore, the SLG-type fault most easily suffers from the negative-sequence dominated instability.

TABLE IV
SIMULATION AND EXPERIMENTAL VERIFICATIONS OF INSTABILITY TYPES AND CURRENT INJECTION LIMITS

| Instability types | Items | SLG ($\dot{I}^- = 0.2\angle 90°$) [a] | DLG ($\dot{I}^- = 0.5\angle 90°$) | LL ($\dot{I}^- = 0.5\angle 90°$) |
|---|---|---|---|---|
| | | (See Figs. 15 and 16) | | |
| Positive-sequence dominated type-1 instability | Equilibrium point analysis | $\dot{I}^+_{lim} = 1.42\angle -30°$ | $\dot{I}^+_{lim} = 0.76\angle -30°$ | $\dot{I}^+_{lim} = 0.94\angle -30°$ |
| | Simulation [b] | $\dot{I}^+_{lim} = 1.41\angle -30°$ | $\dot{I}^+_{lim} = 0.76\angle -30°$ | $\dot{I}^+_{lim} = 0.93\angle -30°$ |
| | Experiment [b] | $\dot{I}^+_{lim} = 1.39\angle -30°$ | $\dot{I}^+_{lim} = 0.73\angle -30°$ | $\dot{I}^+_{lim} = 0.91\angle -30°$ |
| Positive-sequence dominated type-2 instability | Equilibrium point analysis | $\dot{I}^+_{lim} = 1.10\angle 90°$ | $\dot{I}^+_{lim} = 0.59\angle 90°$ | $\dot{I}^+_{lim} = 0.72\angle 90°$ |
| | Simulation [b] | $\dot{I}^+_{lim} = 1.00\angle 90°$ | $\dot{I}^+_{lim} = 0.56\angle 90°$ | $\dot{I}^+_{lim} = 0.67\angle 90°$ |
| | Experiment [b] | $\dot{I}^+_{lim} = 0.94\angle 90°$ | $\dot{I}^+_{lim} = 0.50\angle 90°$ | $\dot{I}^+_{lim} = 0.58\angle 90°$ |
| | | SLG ($\dot{I}^+ = 0.5\angle -90°$) | DLG ($\dot{I}^+ = 0.5\angle -90°$) | LL ($\dot{I}^+ = 0.5\angle -90°$) |
| | | (See Figs. 17 and 18) | | |
| Negative-sequence dominated type-1 instability [b] | Equilibrium point analysis | $\dot{I}^-_{lim} = 0.54\angle -30°$ | $\dot{I}^-_{lim} = 0.92\angle -30°$ | $\dot{I}^-_{lim} = 1.13\angle -30°$ |
| | Simulation [b] | $\dot{I}^-_{lim} = 0.44\angle -30°$ | $\dot{I}^-_{lim} = 0.85\angle -30°$ | $\dot{I}^-_{lim} = 1.04\angle -30°$ |
| | Experiment [b] | $\dot{I}^-_{lim} = 0.46\angle -30°$ | $\dot{I}^-_{lim} = 0.82\angle -30°$ | $\dot{I}^-_{lim} = 1.07\angle -30°$ |
| Negative-sequence dominated type-2 instability [b] | Equilibrium point analysis | $\dot{I}^-_{lim} = 0.41\angle 90°$ | $\dot{I}^-_{lim} = 0.71\angle 90°$ | $\dot{I}^-_{lim} = 0.87\angle 90°$ |
| | Simulation [b] | $\dot{I}^-_{lim} = 0.40\angle 90°$ | $\dot{I}^-_{lim} = 0.66\angle 90°$ | $\dot{I}^-_{lim} = 0.83\angle 90°$ |
| | Experiment [b] | $\dot{I}^-_{lim} = 0.35\angle 90°$ | $\dot{I}^-_{lim} = 0.58\angle 90°$ | $\dot{I}^-_{lim} = 0.72\angle 90°$ |

[a] The amplitude is specified as 0.2 rather than 0.5 because the latter is too large for the SLG-type fault, leading to the negative-sequence dominated instability.
[b] In the simulations and experiments, the current injection limit is determined by gradually increasing the current amplitude in a 0.01 p.u. step.

### C. Impact of the Sequence Coupling on the Stability

When considering the sequence coupling effect, the current injection limit can be identified using a traversal algorithm with the help of numerical computation. The flow chart of the traversal algorithm is given in Fig. 13. Given a current reference in one sequence, the current injection limit in another sequence can be identified using the algorithm.

In Fig. 10(d), the allowable range for the positive-sequence current injection shrinks when injecting a negative-sequence underexcited current at the same time. In Fig. 11(d), on the contrary, the allowable range for the negative-sequence current injection expands when injecting a positive-sequence overexcited current additionally. The vector diagrams depicting the critical operating points are shown in the subfigures (e) and (f). As displayed in Fig. 10(e) and 10(f), the negative-sequence underexcited current injection produces a counteracting effect on $K_1 \vec{U}_g^+ e^{-j\pi/3}$, therefore reducing the positive-sequence current injection limit. In Fig. 11(e) and 11(f), the positive-sequence overexcited current injection provides a supporting effect on $K_4 \vec{U}_g^+ e^{j\pi/3}$, therefore improving the negative-sequence injection limit. Both of these two aspects are attributable to the sequence coupling, and they are also identical with the observation from the phasor diagram in Fig. 9.

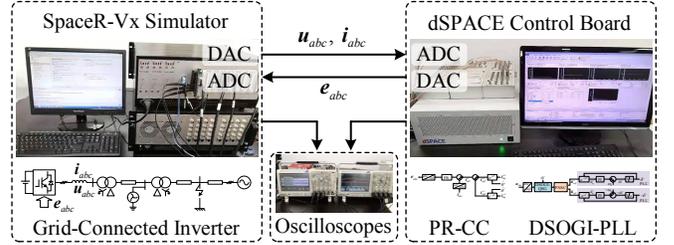

Fig. 14. The HIL experimental setup, where PR-CC refers to proportional-resonant regulator-based current control; ADC and DAC refer to analog-to-digital and digital-to-analog conversions, respectively.

## V. SIMULATION AND EXPERIMENTAL VERIFICATIONS

The instability types and the current injection limit that are defined and identified in Section IV have been verified by simulations and experiments. The simulations are based on an average-value model of inverters, in which the PLL/FLL dynamics, current control dynamics, and network dynamics are modeled. In the experimental verifications, it should be noted that it is improper to emulate asymmetrical grid faults using an unbalanced voltage-source device as it cannot represent the sequence coupling. If a realistic fault case is applied, it will draw large fault currents, which may damage the power electronic inverter in operation. Because of these, a hardware-in-the-loop (HIL) experimental setup is employed, as displayed in Fig. 14. Asymmetrical grid faults are simulated by a SpaceR-Vx real-time simulator, in which the sequence coupling can be represented as expected. In the simulator, a detailed switching model of an inverter is run with a fixed 2e–5 seconds step size. The inverter is controlled by a dSPACE control board with a fixed 1e–4 seconds sampling and control period. The parameters for the simulations and experiments are listed in Table III.

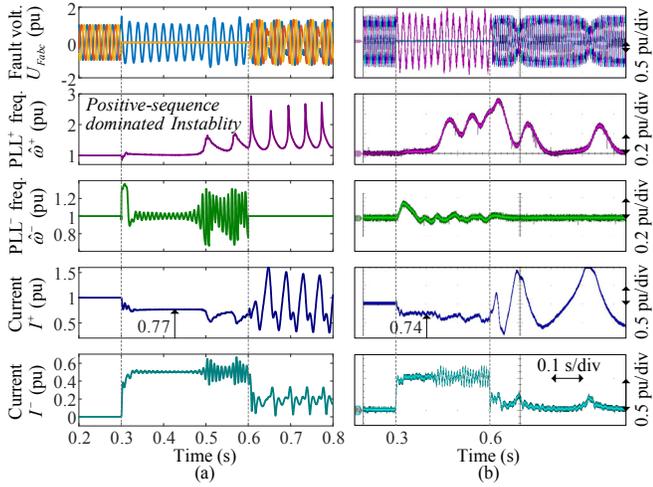

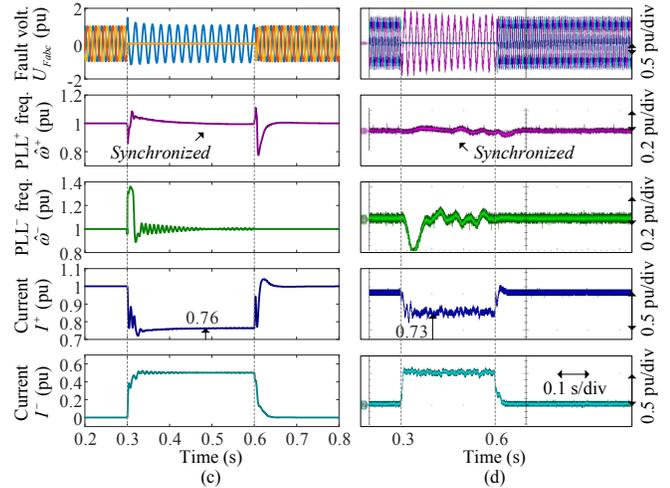

Fig. 15. Positive-sequence dominated type-1 synchronization instability occurs in the simulation (a) $I^+ = 0.77$ and experiment (b) $I^+ = 0.74$. The instability is avoided in the simulation (c) $I^+ = 0.76$ and experiment (d) $I^+ = 0.73$.

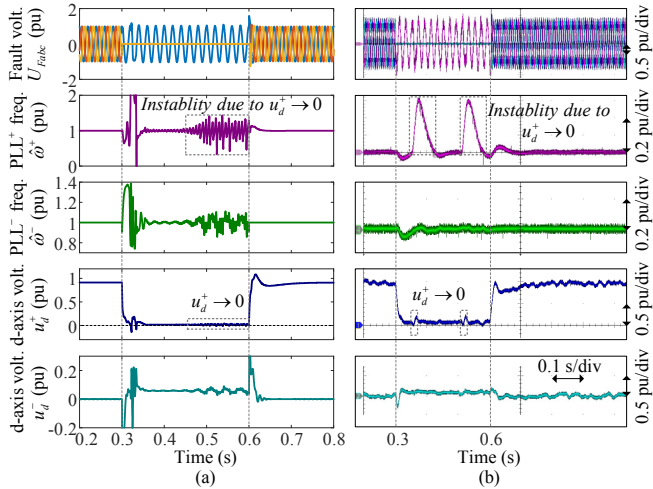

Fig. 16. Positive-sequence dominated type-2 synchronization instability occurs in the simulation (a) $I^+ = 0.57$ and experiment (b) $I^+ = 0.51$ due to $u_d^+ \to 0$. Since $u_d^+$ jumps back and forth near zero, the type-2 instability is mainly manifested as the frequency chattering, which is different from the frequency deviating in the type-1 instability. It has been verified that the instability can be avoided if the current amplitude is reduced by 0.01 p.u.

Considering the current injection limit at $\angle -30°$ and $\angle 90°$ as examples (as exemplified in Figs. 10 and 11), four instability types under three types of asymmetrical faults have been verified. The results are summarized in Table IV. Again, it is noted that the current injection limit identified by the equilibrium point analysis can only reflect the steady-state requirements, and therefore the limit is just a necessary condition for stability. While in simulations and experiments, the dynamic characteristics are considered. Therefore, the current injection limit identified in the simulations and experiments is smaller than that by the equilibrium point analysis. Also, it is seen that instability is more likely to happen in the experiments. This is because the switching harmonics make it easier to reach the current injection limit. In particular, once the *d*-axis voltage decreases to zero, the type-2 instability occurs.

*A. Positive-Sequence Dominated Instability Verifications*

Considering the DLG fault type, it is shown in Fig. 15(a) and (b) that if the current injection exceeds the limit in Table IV, the inverter system will undergo the positive-sequence dominated type-1 instability. In other words, the instability is dominated by the loss of the positive-sequence synchronization. Due to the effect of the sequence coupling, the loss of positive-sequence synchronization leads the negative-sequence control to oscillate. After the fault is cleared, the system is not able to return to normal operation. It should be noted that this instability case is much close to the critical stability, and therefore it takes a relatively long time to accumulate the imbalanced input to force the state of the PLL$^+$ to diverge significantly. If the current reference is notably larger than the limit, it can be verified that the instability will occur rapidly. In an actual power plant/station, the synchronization instability would force the IBG equipment to disconnect [2]. In Fig. 15(c) and (d), it is verified that the instability is avoided, where the amplitude of the positive-sequence current injection is slightly reduced by 0.01 p.u.

In Fig. 16, the simulation and experimental results of the positive-sequence dominated type-2 instability are displayed. It has been indicated that the instability is caused by $u_d^+ \to 0$ because the grid voltage is counterbalanced by the impedance voltage drop completely. The positive-sequence current injection, in this case, is underexcited, which is actually not what the grid codes expect. When injecting an overexcited current, the type-2 instability is unlikely to occur, as demonstrated in Fig. 10(d), which, however, may result in the type-1 instability.

*B. Negative-Sequence Dominated Instability Verifications*

Negative-sequence dominated instability is more likely to occur under the SLG-type fault because the negative-sequence voltage under this type of fault has a small amplitude. Considering the SLG fault type as an example, the negative-sequence dominated instability verifications are conducted. The type-1 instability results can be found in Fig. 17. The instability is manifest as the frequency observed in the negative sequence deviating whereas the one observed in the positive sequence oscillating. Moreover, since the negative-sequence component disappears after the fault is cleared, the system can stabilize. This, however, cannot be always guaranteed as the positive-sequence control may undergo a large oscillation before amplitude of the negative-sequence current injection is reduced by 0.01 p.u., the instability can be avoided.

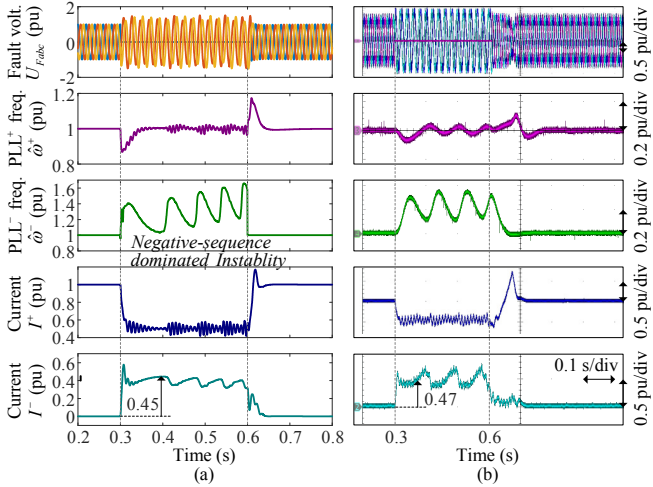

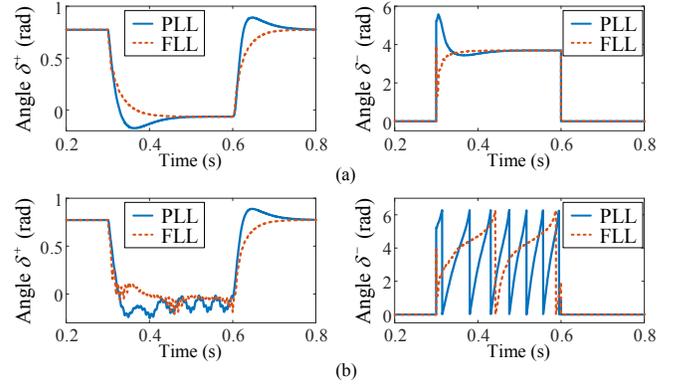

Fig. 19. Comparison of the dynamic response of DSOGI-PLL and -FLL. (a) Stable case. (b) Unstable case, where there is no equilibrium points.

The verifications of the negative-sequence dominated type-2 instability can be found in Fig. 18. As indicated in the dominant factor analysis, the instability is caused by $u_d^- \to 0$ on account of the overmuch negative-sequence underexcited current injection. Theoretically, there is a possibility of such instability, but it can be avoided in practice by applying a dead zone.

### C. Dynamic Response Comparison of DSOGI-PLL and -FLL

Although the dynamic behavior of different PLL/FLL synchronization units may be largely different, synchronization instability will arise no matter which PLL/FLL algorithm is applied, if the steady-state equilibrium point condition cannot be followed. This is the case as shown in Fig. 19, in which it is also seen that PLL and FLL lead to different dynamic responses. Since a PLL generally possesses second-order properties, it shows an overshooting phenomenon in the dynamic response. The overshooting is detrimental to the synchronization stability, as elaborated in [9] for the positive-sequence synchronization. Compared to the positive-sequence synchronization under symmetrical grid faults, it is noted again that there is an interaction between the positive and negative sequences under asymmetrical faults. The sequence coupling is a new issue for the dual-sequence synchronization, which challenges existing analysis approaches for the positive-sequence synchronization stability [7]–[13]. In other words, how to analyze the dual-sequence dynamic synchronization behavior remains unaddressed. In the FLL, the angle is estimated by arctangent functions, and thus its dynamic characteristics are determined mainly by the DSOGI block (RCF/CCF), as indicated in Fig. 7. To understand the dual-sequence dynamic instability behavior (including the effect of the PLL/FLL dynamics, the current loop dynamics, etc.), more analytical research is needed in the future.

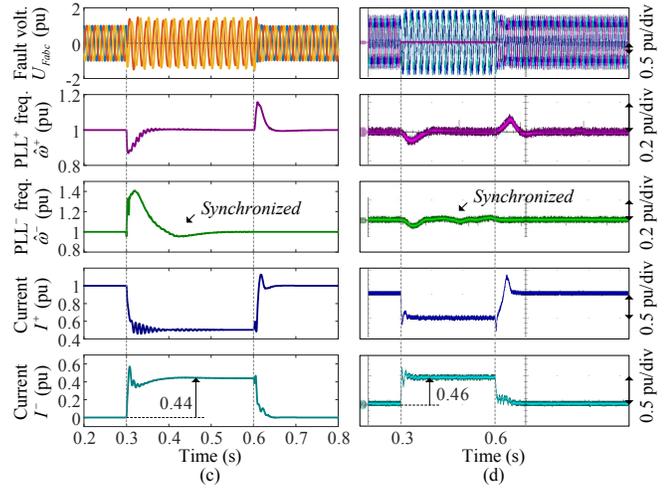

Fig. 17. Negative-sequence dominated type-1 synchronization instability occurs in the simulation (a) $I^- = 0.45$ and experiment (b) $I^- = 0.47$. The instability is avoided in the simulation (c) $I^- = 0.44$ and experiment (d) $I^- = 0.46$.

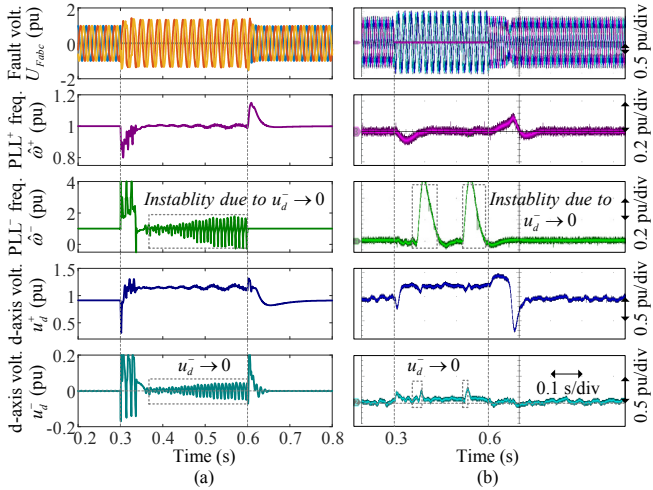

Fig. 18. Negative-sequence dominated type-2 synchronization instability occurs in the simulation (a) $I^- = 0.41$ and experiment (b) $I^- = 0.36$ due to $u_d^- \to 0$. The frequency chattering can be observed in the negative sequence, similar to that in Fig. 16. It has been verified that the instability can be avoided if the current amplitude is reduced by 0.01 p.u.

the fault clearance. In Fig. 17(c) and (d), it is seen that if the

## VI. CONCLUSION

This paper aims to study the dual-sequence synchronization stability of IBG under asymmetrical grid fault conditions. A dual-sequence synchronization model was developed for the quantitative analysis of instability. In the model, the effect of the sequence coupling was appropriately represented by the symmetrical components method. Based on the model, the necessary conditions for the dual-sequence synchronization stability were identified by looking into the existence of equilibrium points. According to the stability conditions, synchronization instability was categorized into four types, each of which was analyzed. The dominant factors in each type of instability were quantified with the current injection limit. This paper provided new insights into describing the dual-sequence synchronization behavior of IBG during asymmetrical faults and quantitatively identifying the important factors affecting the stability.

The findings show that the LOS of a dual-sequence current-controlled IBG signifies that both the positive and negative sequences suffer from instability. The instability, however, is generally dominated by one sequence. The positive-sequence dominated instability is more likely to occur under the DLG fault whereas the positive-sequence dominated one is more likely to occur under the SLG fault. Additionally, the positive-sequence current injection limit is affected by the current injection in the negative sequence and vice versa. The sequence coupling effects caused by the overexcited and underexcited current injection are also opposite. These findings suggest that a tradeoff should be considered in assigning the positive- and negative-sequence reactive currents. This will be an important consideration in the amendments of future grid codes. Based on the findings, future attention will be directed to dynamic synchronization behavior, multi-converter grid-connected system synchronization, etc.

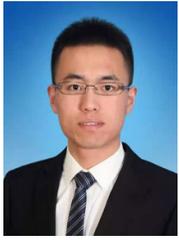

**Xiuqiang He** (S'17–M'21) received the B.S. degree in automation, in 2016, and the Ph.D. degree in control science and engineering, in 2021, from Tsinghua University, Beijing, China.

His current research interests include transient stability and synchronization stability of grid-connected converters and power electronics-dominated power systems.

Dr. He was the recipient of the China National Scholarship, the Beijing Outstanding Graduates Award, the Outstanding Doctoral Dissertation Award of Tsinghua University, the IEEE International Power Electronics and Application Conference and Exposition (PEAC) Excellent Paper Award in 2018, and the IEEE Transactions on Sustainable Energy Outstanding Reviewer Award in 2019.

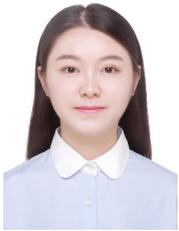

**Changjun He** received the B.S. degree, in 2020, from the Huazhong University of Science and Technology, Wuhan, China. Since 2020, she has been working toward the Ph.D. degree at the Department of Automation, Tsinghua University.

Her current research interests include modeling and analysis of transient stability of the power system with renewable energy generations.

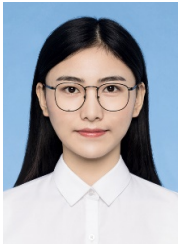

**Sisi Pan** received the B.E. degree, in 2018, from the Department of Electrical Engineering, Yangzhou University, Yangzhou, China, where she is currently working toward the MA.Eng. degree.

Her current research interests include hybrid real-time simulation for dynamic studies of modern power systems.

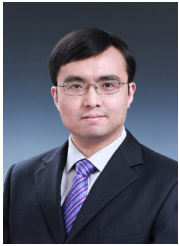

**Hua Geng** (S'07–M'10–SM'14–F'20) received the B.S. degree in electrical engineering from Huazhong University of Science and Technology, Wuhan, China, in 2003 and the Ph.D. degree in control theory and application from Tsinghua University, Beijing, China, in 2008. From 2008 to 2010, he was a Postdoctoral Research Fellow with the Department of Electrical and Computer Engineering, Ryerson University, Toronto, ON, Canada. He joined the Department of Automation Tsinghua University in June 2010 and is currently a full professor.

His current research interests include advanced control on power electronics and renewable energy conversion systems.

Dr. Geng has authored more than 170 technical publications and holds more than 20 issued Chinese/US patents. He was granted the second prize of the National Science and Technology Progress Award. He is the editors of IEEE Transactions on Energy Conversion and IEEE Transactions on Sustainable Energy, associate editors of IEEE Transactions on Industry Applications, IET Renewable Power Generation, Control Engineering Practice. He served as general co-chairs, track chairs, and session chairs of several IEEE conferences. He is an IEEE Fellow and an IET Fellow, convener of the modeling working group in IEC SC 8A, Standing Director of China Power Supply Society (CPSS), vice chair of IEEE IAS Beijing Chapter.

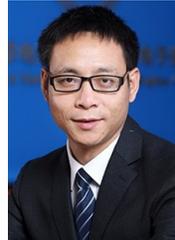

**Feng Liu** (M'10–SM'18) received the B.Sc. and Ph.D. degrees in electrical engineering from Tsinghua University, Beijing, China, in 1999 and 2004, respectively. He is currently an Associate Professor with Tsinghua University. From 2015 to 2016, he was a Visiting Associate with California Institute of Technology, Pasadena, CA, USA.

His research interests include stability analysis, optimal control, robust dispatch and game theory based decision making in energy and power systems.

Dr. Liu is the author/coauthor of more than 200 peer-reviewed technical papers and two books, and holds more than 20 issued/pending patents. He is an Associated Editor of several international journals, including the IEEE Transactions on Smart Grid, IEEE Power Engineering Letters, IEEE Access, etc. He also served as a Guest Editor for the IEEE Transactions on Energy Conversion.